# Writing With Machines and Peers: Designing for Critical Engagement with Generative AI

Xinran Zhu*, Cong Wang, Duane Searsmith
University of Illinois Urban-Champaign

**Abstract**

The growing integration of generative AI in higher education is transforming how students write, learn, and engage with knowledge. As AI tools become more integrated into classrooms, there is an urgent need for pedagogical approaches that help students use them critically and reflectively. This study proposes a pedagogical design that integrates AI and peer feedback in a graduate-level academic writing activity. Over eight weeks, students developed literature review projects through multiple writing and revision stages, receiving feedback from both a custom-built AI reviewer and human peers. We examine two questions: (1) How did students interact with and incorporate AI and peer feedback during the writing process? and (2) How did they reflect on and build relationships with both human and AI reviewers? Data sources include student writing artifacts, AI and peer feedback, AI chat logs, and student reflections. Findings show that students engaged differently with each feedback source—relying on AI for rubric alignment and surface-level edits, and on peer feedback for conceptual development and disciplinary relevance. Reflections revealed evolving relationships with AI, characterized by increasing confidence, strategic use, and critical awareness of its limitations. The pedagogical design supported writing development, AI literacy, and disciplinary understanding. This study offers a scalable pedagogical model for integrating AI into writing instruction and contributes insights for system-level approaches to fostering meaningful human–AI collaboration in higher education.

**Keywords:** AIED, academic writing, hybrid intelligence, peer feedback, human-AI collaboration

## Introduction

The rapid advancement of generative artificial intelligence (GenAI) is reshaping how people learn, think, and create knowledge. In education, GenAI offers new possibilities across a wide range of activities—from automating routine tasks such as grammar correction to supporting more complex processes like generating lesson plans and engaging in collaborative knowledge building (Chen & Zhu, 2023; Moundridou et al., 2024). As AI tools become increasingly embedded in various sectors of society, a key challenge lies in ensuring that humans are equipped with the skills needed to interact with these systems meaningfully, critically, and reflectively. This requires not only an understanding of AI's capabilities and limitations, but also the ability to navigate the often opaque and ambiguous situations that result from the lack of transparency in how many AI systems operate (Bearman & Ajjawi, 2023; Long & Magerko, 2020; Shibani et al., 2024).

This challenge is particularly pressing in education, where effective AI use depends on pedagogically grounded and domain-specific integration. Without the necessary skills, students may use AI in superficial or even harmful ways—for instance, bypassing critical thinking, reinforcing misconceptions, or amplifying bias and misinformation (Fan et al., 2025; Chen et al., 2025). Developing students' critical engagement with AI is therefore an urgent priority for both technological design and pedagogical practice (Shibani et al., 2022). Such engagement requires students to think critically and reflectively about AI outputs, recognizing their limitations and making informed decisions about when and how to use them (Shibani et al., 2024).

In the context of academic writing, there is growing attention to the role of AI across different stages of the writing process. Writing is a complex, recursive, and socially situated activity that unfolds through iterative cycles of drafting, feedback, and revision (Flower & Hayes, 1981; MacArthur et al., 2015). To support this process, a range of AI-supported tools have been introduced, including conversational systems like ChatGPT (Barrett & Pack, 2023), analytic feedback platforms such as AcaWriter (Knight et al., 2020), and co-authoring tools that facilitate human–AI collaboration (Lee et al., 2022). While these tools show promise in enhancing both argumentative and creative writing, their integration into classroom practice raises important questions about how students actually engage with them in authentic learning environments. Despite growing interest, empirical evidence from classrooms is still limited (Shibani et al., 2024). Specifically, it remains unclear about how students navigate writing tasks that span multiple stages and platforms, how they integrate AI-generated and human feedback in dynamic and iterative ways, and how they build critical partnerships with AI by leveraging the complementary strengths of human judgment and machine support throughout the writing process.

* Corresponding author. Contact: xrzhu@illinois.edu

Building on this emerging line of work, our study designed and investigated a pedagogical approach that integrates GenAI-generated and peer feedback within an authentic academic writing context. We conducted the study in a graduate-level course where students developed literature review projects through multiple iterative revisions. At key stages, students received feedback from a custom GenAI reviewer, participated in structured peer review and class discussions, and reflected on both types of input. We ask two research questions: (1) How do students interact with and incorporate AI and peer feedback during the writing process? and (2) How do they reflect on and build relationships with both AI and human reviewers? By tracing revision trajectories, analyzing interaction patterns, and examining student reflections, we unpack the social-cognitive processes through which learners engage with a hybrid feedback system that combines peer and AI input.

Situated in an authentic classroom setting, this study provides nuanced insights into how students make sense of, respond to, and learn from both AI and peer feedback. It also examines how students build reflective and critical relationships with AI tools throughout the writing process. In doing so, the study contributes to our understanding of human–AI collaboration and offers a pedagogical framework to inform broader, system-level implementations of AI-supported writing instruction.

## Related Work

### Academic Writing and Peer Feedback

Writing is a complex and recursive process that unfolds through iterative cycles of drafting, feedback, and revision (Flower & Hayes, 1981). Writing is also a socially situated activity. It involves using language in structured ways to make meaning within specific contexts, while responding to, influencing, and being influenced by others (Cope & Kalantzis, 1993). Writers engage in this social exchange to reflect on their thinking, construct meaning, develop ideas, and express identities in response to evolving social contexts (Flower & Hayes, 1981; MacArthur et al., 2015). At the same time, writing serves as a knowledge-building practice. Expert writers engage in knowledge-transforming processes, using revision not just to refine language, but to rethink goals and restructure ideas (Bereiter & Scardamalia, 1987).

In academic writing, feedback plays a central role in the revision process. Traditionally provided by teachers or peers, feedback offers information about one's performance or understanding, guiding writers toward improvement of their work (Hattie & Timperley, 2007). Feedback functions not only as an evaluative tool but also as a mechanism for enhancing writing performance, fostering self-regulation, and supporting knowledge practices (Nicol & Macfarlane-Dick, 2006; Tan & Chen, 2022; Yang et al., 2016). Within socio-cultural and academic literacies frameworks, feedback is further conceptualized as a dialogic and participatory process that enables students to engage with diverse perspectives, negotiate meaning, and co-construct disciplinary knowledge (Carless & Boud, 2018; Sutton, 2012). Peer feedback, in particular, serves as a form of collaborative learning, where students take on the dual roles of giving and receiving feedback to deepen their reflection and promote active participation in the writing process (Kollar & Fischer, 2010). Given its central role in learning, there is growing interest in how to deliver feedback more effectively—especially in ways that provide constructive and actionable guidance that leads to the incorporation of feedback (Fong et al., 2021; Tan & Chen, 2022).

Despite its importance for writing and learning, feedback doesn't always lead to improvement. Not all feedback can—or should—be incorporated, and its impact depends largely on how students understand, value, and act on it (Carless & Boud, 2018). Prior literature has identified several factors that influence feedback uptake. First, the quality of feedback is essential. For peer feedback to be effective, reviewers must possess sufficient knowledge to provide relevant and constructive input (Cho et al., 2006). Moreover, feedback that is overly complex, poorly structured, or assumes background knowledge that students do not yet possess can also be difficult to interpret and apply (Nelson & Schunn, 2009). Second, the *specificity* of feedback plays a key role in shaping student responses. Feedback that offers clear, concrete suggestions or solutions is more likely to be taken up, as it provides actionable guidance that students can readily understand and implement (Tan & Chen, 2022; Nelson & Schunn, 2009). In contrast, vague or general comments that merely describe issues without identifying specific problems or recommending next steps may leave students unsure how to proceed, thereby limiting meaningful engagement (Chen et al., 2024; Nelson & Schunn, 2009). Finally, learner-related factors such as self-regulation, self-efficacy, and emotional responses also affect uptake. Students with weaker self-regulatory skills may struggle to process and act on feedback effectively (Nelson & Schunn, 2009).

With the rise of digital tools (Kruse et al., 2023), writing assistance technologies have started to take on roles traditionally filled by human reviewers—such as diagnosing writing problems, offering targeted feedback, and guiding revision—thereby reshaping how feedback is delivered and used in academic writing. In addition to automated grammar checkers or summative scoring systems, intelligent writing assistants are increasingly positioned as interactive collaborators in the writing process (Lee et al., 2024). For example, AcaWriter guides students to attend to the rhetorical structure of academic writing by visually highlighting features such as argumentation and evidence, thereby prompting more conscious and informed revision (Knight et al., 2020). Similarly, CyberScholar integrates a rubric-based AI chatbot that provides evaluative feedback and supports dialogue-based reflection, allowing students to receive targeted suggestions and engage in interactive feedback exchanges within a writing task (Castro et al., 2025).

The growing integration of digital tools into writing feedback processes—especially with the recent rise of GenAI technologies—invites a broader reconceptualization of human–AI collaboration. This shift raises important questions about how to support students' critical engagement with AI tools and how human and AI feedback can work together to enhance writing and learning (Shibani et al., 2024).

**Promoting Critical Engagement with AI in Academic Writing Feedback**

Shibani et al. (2022) define critical engagement as "the act of questioning engagement with data, analytics and computational tools with an understanding of its limitations and assumptions, alongside the analytical ability and agency to challenge its outcomes when necessary." Such engagement involves metacognitive and reflective capacities that enable learners to question, interpret, and selectively incorporate AI-generated input (Shibani et al., 2024). In the context of academic writing, Shibani et al. (2024) argue that fostering such "critical interaction" skills is essential for students to use AI feedback meaningfully. However, their study shows that students often engage with AI in a superficial manner, with limited evidence of reflective uptake. Without adequate scaffolding, learners may rely too heavily on AI outputs, potentially bypassing deeper learning. This concern mirrors findings from peer feedback research, which shows that feedback effectiveness depends not only on its quality but also on students' ability to interpret and apply it (Schunn et al., 2016).

Engaging critically with AI requires that learners understand both what AI can do and where its limits lie. It also involves recognizing the complementary roles that humans and machines can play in learning environments. The recently emerging concept of hybrid intelligence (HI) provides a helpful lens for thinking about this collaboration. HI refers to the synergistic integration of human and artificial cognitive systems to achieve outcomes that neither could attain independently (Akata et al., 2020; Dellermann et al., 2019). Akata et al. (2020) emphasize that such collaboration enables humans to address complex, open-ended problems by combining the computational efficiency of artificial intelligence (AI) with human contextual judgment, creativity, and ethical reasoning. HI is grounded in an understanding of the distinct affordances and limitations of both human and artificial cognition. Human cognition is adaptive and situated, shaped by lived experiences, social interactions, and cultural contexts. It is marked by creativity, self-reflection, and the capacity for moral and pedagogical reasoning (Siemens et al., 2022). In contrast, artificial cognition operates through algorithmic processing and pattern recognition. It is capable of executing tasks with remarkable speed, scale, and consistency, yet lacks emotional intelligence, contextual sensitivity, and the ability to generalize across novel domains (Banihashem et al., 2023; Korteling et al., 2021).

This understanding of HI has important implications for writing and feedback to understand both human and AI contributions' distinct but complementary affordances. Empirical studies demonstrate that AI-generated feedback offers timely, structured, and rubric-aligned guidance that supports surface-level improvements in grammar, coherence, and formatting (Li et al., 2024; Steiss et al., 2024). However, AI frequently falls short in providing contextually aware and pedagogically meaningful suggestions for higher-order aspects such as argumentation, rhetorical clarity, and conceptual development (Steiss et al., 2024; Solovey, 2024). Human feedback, by comparison, is more nuanced and responsive to students' disciplinary and developmental needs, though it is less scalable and slower to deliver (Solovey, 2024; Steiss et al., 2024). Recognizing these complementary affordances, recent studies have begun to design hybrid feedback systems that intentionally integrate both human and AI input. For example, Suraworachet et al. (2023) implemented a hybrid feedback intervention in which students received tutor-written comments on the content and quality of their weekly reflective writing, alongside AI-generated feedback based on Google Docs activity logs. The AI feedback focused on students' writing engagement patterns, such as the frequency and regularity of their writing, thus supporting both cognitive and behavioral dimensions of the task.

For such hybrid feedback systems to be effective in promoting feedback uptake and improving learning outcomes, intentional pedagogical design is essential. Banihashem et al. (2025) propose a hybrid intelligent feedback framework that conceptualizes how humans and GenAI can collaborate across a spectrum of roles in the feedback process—from fully human-led to fully AI-led—depending on task complexity, instructional intent, and desired trade-offs between efficiency and pedagogical richness. Their framework also outlines a six-phase pedagogical design process to support this collaboration: defining feedback purposes, aligning with learner needs, selecting appropriate hybrid modes, generating feedback, evaluating its quality, and delivering it in ways that foster shared agency.

Building on these conceptual and empirical foundations, this study investigates how a hybrid feedback system—integrating both GenAI and peer feedback—can support academic writing in higher education. Although prior research has underscored the complementary strengths of AI and human feedback, as well as the importance of developing critical interaction skills, there remains limited understanding of how students actually engage with and make sense of hybrid feedback in authentic learning environments. To address this gap, we designed and implemented a pedagogical framework that integrated GenAI and peer feedback throughout a multi-stage, iterative writing task in an eight-week graduate-level course. This framework provided opportunities for students to engage purposefully with both AI and peer reviewers at key points in the writing process. By analyzing feedback interaction patterns, tracing revision trajectories, and examining student reflections, we explore how students engaged with hybrid feedback and how their relationships with AI tools evolved over time.

## Methods

To answer the research questions, we employed a case study–mixed methods design, in which a parent case study incorporated a nested mixed methods approach (Guetterman & Fetters, 2018). As a methodology, case study allows us to develop an in-depth understanding of the nuance of the design enactment on a manageable scale (Stake, 1978), while also retaining the potential to extract ideas at a more abstract level from the case study that can pertain to other contexts (Yin, 2009, Yin, 2013). The case in this study was defined as a graduate-level writing class in which students engaged with AI and peer feedback as part of a multi-stage writing process. Individual students served as embedded units of analysis within the single case (Yin, 2014), allowing us to examine both broad patterns and specific trajectories of engagement. Mixed methods enriched the inquiry by combining and integrating qualitative and quantitative analyses to develop a more comprehensive understanding of students' experiences (Creswell & Plano Clark, 2011). The overall design aimed to generate insight into how students interacted with and incorporated feedback, as well as how they reflected on and built relationships with both human and AI reviewers. In the sections that follow, we describe the research context and participants, the pedagogical and technological designs, and our data sources and analytical approach.

**Context and Participants**

This study was conducted in the context of an 8-week graduate-level course in educational technologies, offered in Spring 2025 at a large research university in the United States. Twelve students enrolled in this class, most of whom were doctoral students in the social sciences, including education, communication, and liberal arts. Ten students consented to participate in this study. The class explored both theoretical and practical applications of emerging learning technologies, such as game-based learning, learning analytics, and artificial intelligence. It was delivered through *CGScholar*, a community-based learning management system (LMS) designed to support student interaction, peer feedback, and collaborative learning. All course content, discussions, and writing activities were hosted on this platform.

As a final assignment, students were required to write a 3500-word paper reviewing the theoretical foundations and practical applications of a learning technology discussed in the class. Students could choose between two project formats: a literature review or a landscape review. The literature review option asked students to frame clear research questions and synthesize academic literature to identify knowledge gaps and explore real-world problems in their fields. The landscape review invited students to examine the use of an emerging technology in a specific educational context, producing a practitioner-oriented analysis. Regardless of format, the assignment emphasized a critical review of relevant theories, current practices, implementation challenges, and potential solutions. Students were asked to connect educational theory with practical application, consider multiple perspectives, and situate their analysis within their own academic and professional interests. The assignment served as a synthesis of course learning, with flexibility for students to pursue topics aligned with their backgrounds and goals.

**Pedagogical and Technological Designs**

*Pedagogical Design*

To support students in completing their final writing project, the course incorporated intentionally designed scaffolds throughout the semester (see Figure 1). Weeks 1-4 focused on collaborative idea development, facilitated through weekly synchronous discussions. During these sessions, students worked in groups of three to share emerging ideas, offer feedback, and engage in brainstorming. In Week 4, students submitted their initial drafts (v1) for feedback from a custom-designed AI tool, *CyberScholar* (see next section for details), which was integrated into the class LMS. The system allowed students to submit drafts for automated feedback, receive rubric-aligned evaluations, and engage in follow-up dialogue with the AI reviewer. Students were able to revise and resubmit their drafts for multiple rounds of AI review at any point.

**Figure 1**
*Writing Activity Design*

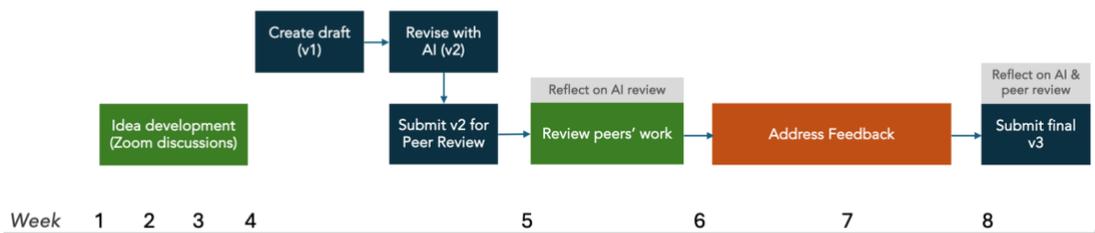

The rubrics are usually developed by subject area experts. In the context of this study, the students used a rubric developed by a group of researchers in literacy and learning technologies, named "Ways of Knowing: Frontier Research" (see Appendix A). This rubric assesses work through the Learning by Design framework (Cope & Kalantzis, 2015), focusing on knowledge processes—experiencing, conceptualizing, analyzing, and applying—plus communication and referencing. Each dimension is rated on a 0–4 scale with descriptive performance levels. It measures how deeply students engage with ideas and how clearly they present them.

Following the AI review, students revised their drafts (v2) and submitted them in Week 5 for structured peer review within the course LMS. Peer review included both in-text annotations and holistic evaluations using the same rubric as the AI system. Each student was asked to provide reviews to 2 peers in their groups. These reviews were further discussed during in-class peer feedback sessions, where students met with their group members to clarify and discuss suggestions with their reviewers. Final drafts (v3) were submitted at the end of the course in Week 8.

To promote critical reflection on the writing and revision process, students completed two written reflections: one after the AI review and one after the peer review. These reflections documented their interactions with each feedback source, described how they engaged with suggestions, and discussed their evolving relationship with AI and peer feedback.

*Technological Design - CyberScholar*

CyberScholar is an AI-driven learning platform designed to support scholarly writing and feedback processes in educational settings. Built as an extension of the CGScholar LMS, it integrates GenAI capabilities to provide automated, rubric-aligned feedback while fostering critical human-AI collaboration. The platform emphasizes "cyber feedback"—AI reviewer that amplifies human cognitive processes rather than replacing them—aligning with hybrid intelligence principles (Akata et al., 2020). In this study, CyberScholar served as the GenAI reviewer, enabling students to submit drafts, receive detailed evaluations, and engage in iterative dialogues to refine their work.

At its core, CyberScholar employs a modular architecture that combines frontend and backend technologies for seamless user interaction. The frontend offers a responsive interface where students can upload drafts via a rich text editor that supports multimodal content, including text, images, and embedded media. The backend handles data storage using a document store for user works and a relational database for user profiles, ensuring secure and scalable operations.

The primary feedback mechanism, *CyberReview*, leverages GenAI models accessed through an AI model routing framework, including GPT-4o (used by most students in this study). Feedback is generated against custom rubrics, such as the "Ways of Knowing: Frontier Research" rubric used in this study. This rubric evaluates criteria like theoretical framing, evidence integration, and innovation on a 0-4 scale, with descriptive performance levels.

To enhance contextual relevance, CyberReview incorporates a Retrieval-Augmented Generation (RAG) system. Previous student work and relevant publications (e.g., on AI in education and writing analytics) are parsed into approximately 300-word chunks and embedded into a Pinecone vector database using semantic embeddings. For each rubric criterion, the system generates a query vector from the submitted draft's content. It then retrieves a configurable number of top-matching chunks (typically 7-10) from the vector store, which are injected into the AI prompt as additional context. This RAG approach ensures feedback is grounded in domain-specific knowledge, while adapting to the draft's unique focus. The AI then produces a structured response per criterion, including a score, justification, and actionable suggestions.

Students interact with CyberReview through a conversational interface (see Figure 2 for a screenshot example), where they can submit drafts, view aggregated feedback, and engage in follow-up chats. For instance, users might query clarifications ("How can I improve my theoretical framing?") or request examples, prompting the AI to generate responses that encourage reflection rather than direct content creation. This dialogic design promotes critical engagement, as students must evaluate and adapt AI outputs, addressing limitations like generic suggestions by leveraging RAG for specificity.

**Figure 2**
*Screenshot of CyberScholar interface showing rubric-aligned feedback and chat dialogue.*

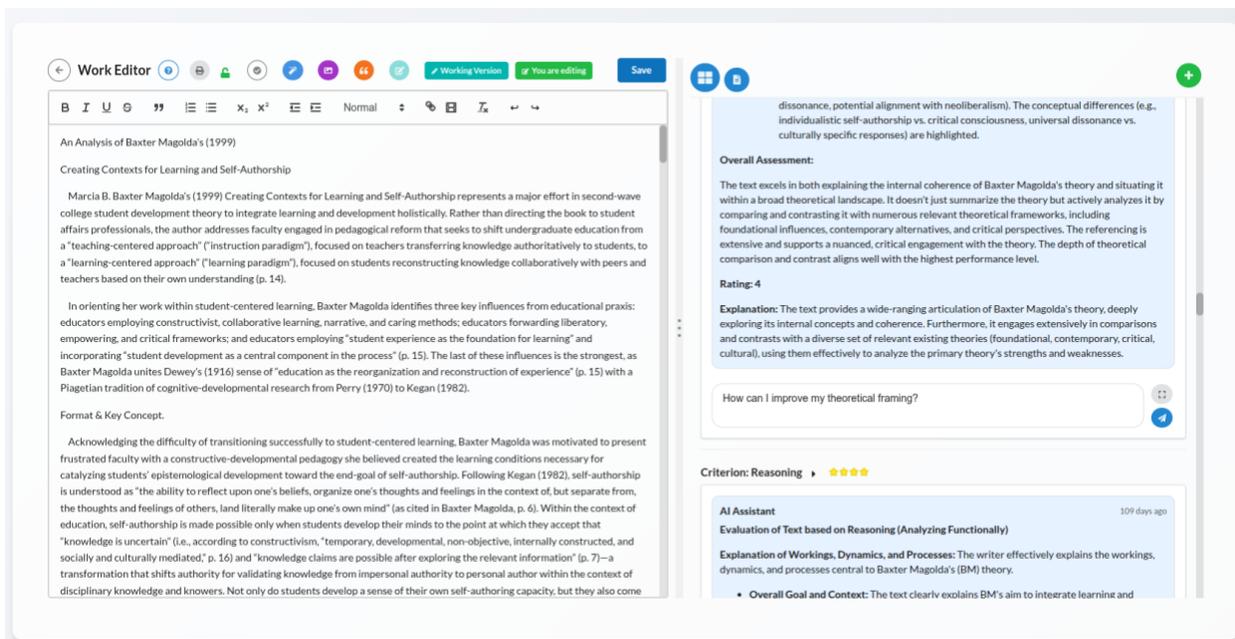

**Data and Analytical Approach**

Following a case study–mixed methods design, we collected and analyzed both quantitative and qualitative data drawn from multiple sources. Quantitative data included log data from CyberScholar, capturing students' interactions with the AI tool. Qualitative data included three versions of each student's writing, written reflections, which provided insight into their revision processes and feedback incorporation. Reviews from peers were also collected as secondary data sources to contextualize how students interpreted and responded to peer feedback. These data allowed us to examine patterns of engagement across the whole class and to explore individual students' revision trajectories in depth. Two students didn't complete the second reflection assignment, so their data for that activity were excluded from the analysis.

For the first research question—how students interacted with and incorporated AI and peer feedback in the writing process—we applied a mixed methods approach. We began with a descriptive analysis of CyberScholar log data, student reflections, and peer reviews, focusing on their engagement patterns with both AI and peer review. We analyzed students' interaction patterns with AI by examining their chat histories and written reflections, using a coding scheme adapted from Shibani et al. (2024), which focuses on critical interaction with AI in writing contexts (see Appendix B). The adapted scheme includes four dimensions of student engagement: planning and ideation, information seeking and evaluation, writing and presentation, and conversational engagement. Each dimension was coded as deep, shallow, or absent to capture the level of student engagement. To examine student revision patterns, we conducted automated text analysis on three versions of each student's writing. Specifically, we used bigram-based n-gram analysis to identify patterns of textual addition, deletion, and lexical similarity across drafts, allowing us to capture surface-level and structural changes over time.

Based on the combined findings from both quantitative and qualitative analyses, we identified two students whose engagement trajectories were representative of distinct interaction patterns with AI and peer feedback. To explore these cases in greater depth, we conducted thematic analysis (Braun & Clarke, 2006) of all related artifacts, including their writing drafts, written reflections, peer feedback, and CyberScholar log data. Thematic coding followed an inductive approach to identify emerging themes related to feedback uptake, revision decisions, and engagement with AI. This approach allowed us to unpack the nuanced ways these students engaged with hybrid feedback across the writing process.

For the second research question—how students reflected on their interactions and built relationships with AI and peer reviewers throughout the writing process—we conducted descriptive coding of students' written reflections (Miles et al., 2019). Using an inductive approach, we coded the reflections to identify themes related to students' understanding of the strengths and limitations of both AI and peer feedback, as well as their evolving relationships with AI. This qualitative analysis complemented the findings from RQ1 by providing insight into how students made sense of their collaborative experiences with both AI and peers, and how they developed critical and reflective relationships with the AI tool across the writing process.

## Results

**How did students engage and incorporate AI and peer feedback in the writing process?**

To examine how students engaged with AI and peer feedback, we conducted a descriptive analysis using three primary data sources: (1) CyberScholar log data, (2) students' writing revisions, and (3) students' written reflections. These sources provided insight into students' interaction patterns with different feedback sources and their revision decisions across the overall writing processes. As a secondary data source, peer-generated reviews were used for triangulating findings to support the interpretation of patterns identified in the main data sources. In the following sections, we first present an overview of students' interaction patterns and revision trajectories, followed by in-depth analysis of two cases—one illustrating critical engagement and sophisticated revisions, the other demonstrating surface-level interaction and uptake.

### *Descriptive Overview of Engagement Patterns*

To reiterate, students submitted their first draft (v1) for AI review in Week 4 via the CyberScholar AI feedback platform. After revising their work based on AI feedback, they submitted a second draft (v2) for peer review, also in Week 4. Following peer feedback, students revised again and submitted their final version (v3) in Week 8. Drafts v2 and v3 were completed through the course LMS, CGScholar.

**Engagement with AI review.** Students exhibited varied levels of interaction with the AI tool. As shown in Table 1, log data from CyberScholar indicated that five students (e.g., Riley) completed only one review session—submitting a full draft and receiving automated feedback once. In contrast, another five students (e.g., Taylor) engaged in multiple sessions across different stages of writing, using the tool iteratively. For instance, Morgan submitted outlines and early drafts over four sessions, gradually developing her work with AI support. In her reflection, she described using the AI to "check [her] work and see the suggestions it has for [her] at this stage." She continued using AI feedback to monitor progress across rubric criteria and to check whether her revisions aligned with expectations. Engagement depth also varied: seven students had minimal interaction with the AI (fewer than four messages sent). For example, Riley submitted her draft for review but did not send any follow-up messages. In contrast, three students engaged more extensively—Taylor exchanged 15 messages, Ethan 17, and Morgan 14. Most students used the ChatGPT-4o model, with one using LLaMA-3.3-70B-Instruct. Notably, Sydney switched between different models across sessions.

**Table 1**
*Student Interaction with CyberScholar*

| Student | Num. of AI Review Sessions | Num. of Student-Initiated Messages | AI Model Used |
|---|---|---|---|
| Riley | 1 | 0 | chatgpt-4o |
| Taylor | 2 | 15 | chatgpt-4o |
| Ethan | 1 | 17 | chatgpt-4o |
| Morgan | 4 | 14 | chatgpt-4o |
| Casey | 3 | 2 | chatgpt-4o |
| Emma | 1 | 0 | chatgpt-4o |

| Alex | 2 | 0 | chatgpt-4o |
| Alice | 1 | 3 | chatgpt-4o |
| Sydney | 2 | 5 | Chatgpt-4o & llama-3.3-70b-instruct |
| Leo | 1 | 1 | llama-3.3-70b-instruct |

*Note.* Student names are pseudonyms.

**Figure 3**
*Student Engagement Level with the CyberScholar AI Tool*

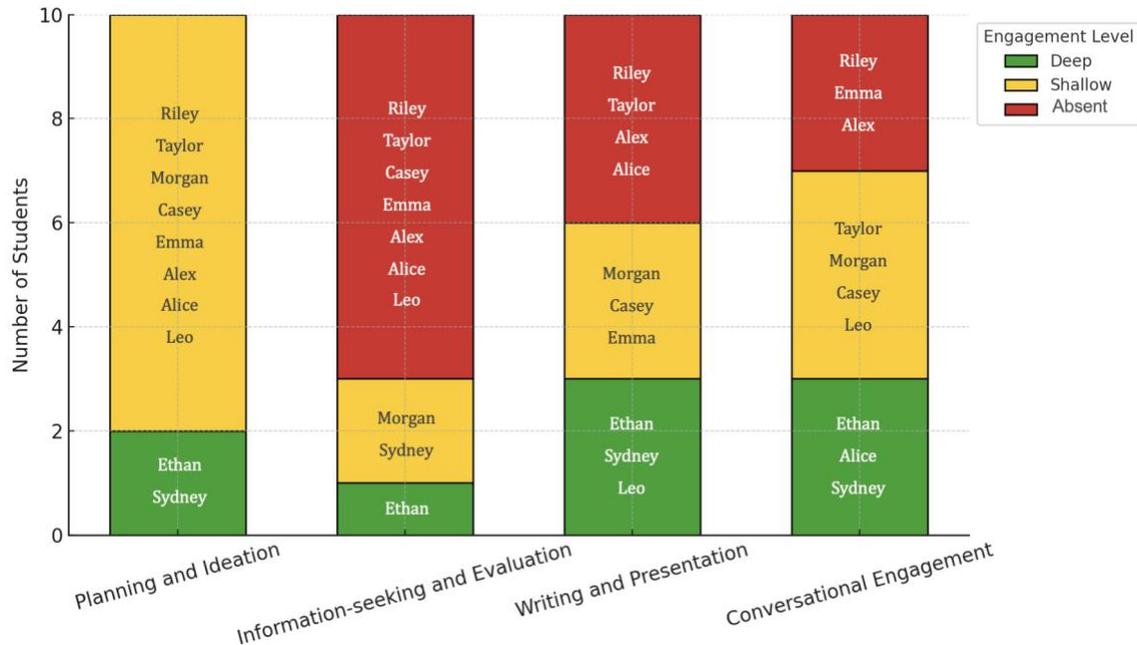

Figure 3 provides an in-depth overview of students' engagement levels with the AI tool across four dimensions. Overall, most students exhibited no (red) or shallow (yellow) engagement, with only a few demonstrating deep engagement (green). Deep engagement was characterized by purposeful use of AI feedback for planning revisions, generating ideas, seeking and evaluating information critically, improving writing flow and coherence, and engaging in dialogic, interactive conversations with the AI. Notably, Ethan demonstrated deep engagement across all four dimensions. In his chat with the AI, he engaged in practices such as asking clarification questions, negotiating interpretations, requesting and evaluating resources, and deliberating on revision decisions. In contrast, Morgan remained shallow in all dimensions—despite both she and Ethan having the highest number of messages exchanged with the AI tool. Similarly, Taylor also had a high message count (14), but like Morgan, her approach was more task oriented. She primarily asked the AI to summarize feedback under each rubric criterion, especially when the initial feedback was lengthy. Her interactions were frequent but remained transactional rather than exploratory or dialogic. This contrast highlights that a high volume of interaction does not necessarily reflect meaningful or deep engagement.

Students with shallow or no conversational engagement tended to treat the AI tool as a rubric-based scoring engine. Chat histories and reflections showed that some students—such as Casey, and Leo—engaged in brief, one-shot exchanges with the chatbot, typically posing a single prompt or question without any follow-up. These students primarily used the tool to quickly identify issues in their work and implement surface-level suggestions. Their interactions largely relied on the scores generated by the AI; lower scores tended to receive more attention, while higher scores were often interpreted as indicating no need for further improvement. Emma, for example, explained: "Since the AI gave my work a four-star rating across all criteria except for Communication (for which I received three stars), most of the feedback I incorporated involved grammatical corrections." This pattern suggests a score-driven approach, where students used the AI tool primarily to target specific areas for improvement in pursuit of a higher rating, rather than engaging in deeper, reflective revisions.

**Engagement with peer review.** Following the AI review and initial revisions, students received peer feedback consisting of both rubric-based scores with justifications and detailed in-text annotations—on average, about 41 comments per student from three peers. Although replies or interactions with comments on the platform were minimal (e.g., replies to peers' comments), students' written reflections revealed thoughtful strategies for managing and acting on this feedback.

Across reflections, students demonstrated intentional approaches to organizing and prioritizing peer comments. These strategies varied in complexity but could be grouped into three overarching types: (1) *Initial screening strategies*, where students read through all comments to gain a holistic understanding of peer impressions before acting—for example, Emma and Morgan emphasized reviewing overall feedback to identify key concerns before addressing specific points; (2) *Organization and tracking strategies*, which involved using external tools or categorization to make feedback more manageable—Taylor and Alex copied annotations into Word documents, with Taylor marking comments by complexity, while Ethan categorized feedback into content-focused, language-focused, and formatting-focused types; (3) *Prioritization strategies,* where students synthesized feedback across reviewers to detect recurring patterns, such as vague arguments or missing citations, and addressed those first. Others, like Alice and Alex, began with simpler tasks—like grammar and formatting edits—to build momentum for more substantive revisions; and (4) *Leveraging live peer discussions*, where students found real-time exchanges more meaningful and easier to process than written comments. For example, Tiara focused more on feedback received during breakout room discussions, and Casey noted that Zoom interactions provided richer context and a more encouraging tone. Collectively, these patterns suggest that students were not merely responding to feedback passively, but actively transforming it into structured, actionable revision plans aligned with their goals.

*Descriptive Overview of Revision Trajectories*

To understand students' revision patterns across different feedback stages, we first generated n-grams (n = 2) from each draft and calculated a set of automated textual similarity and change metrics between drafts (Figure 3). These included Jaccard similarity to assess overall lexical overlap, and counts of added and deleted n-grams to reflect surface-level textual edits. These quantitative measures aimed to provide an overview of revision behaviors, which we then contextualized through qualitative analysis of student artifacts and reflections.

**Figure 3**
*Revision Metrics*

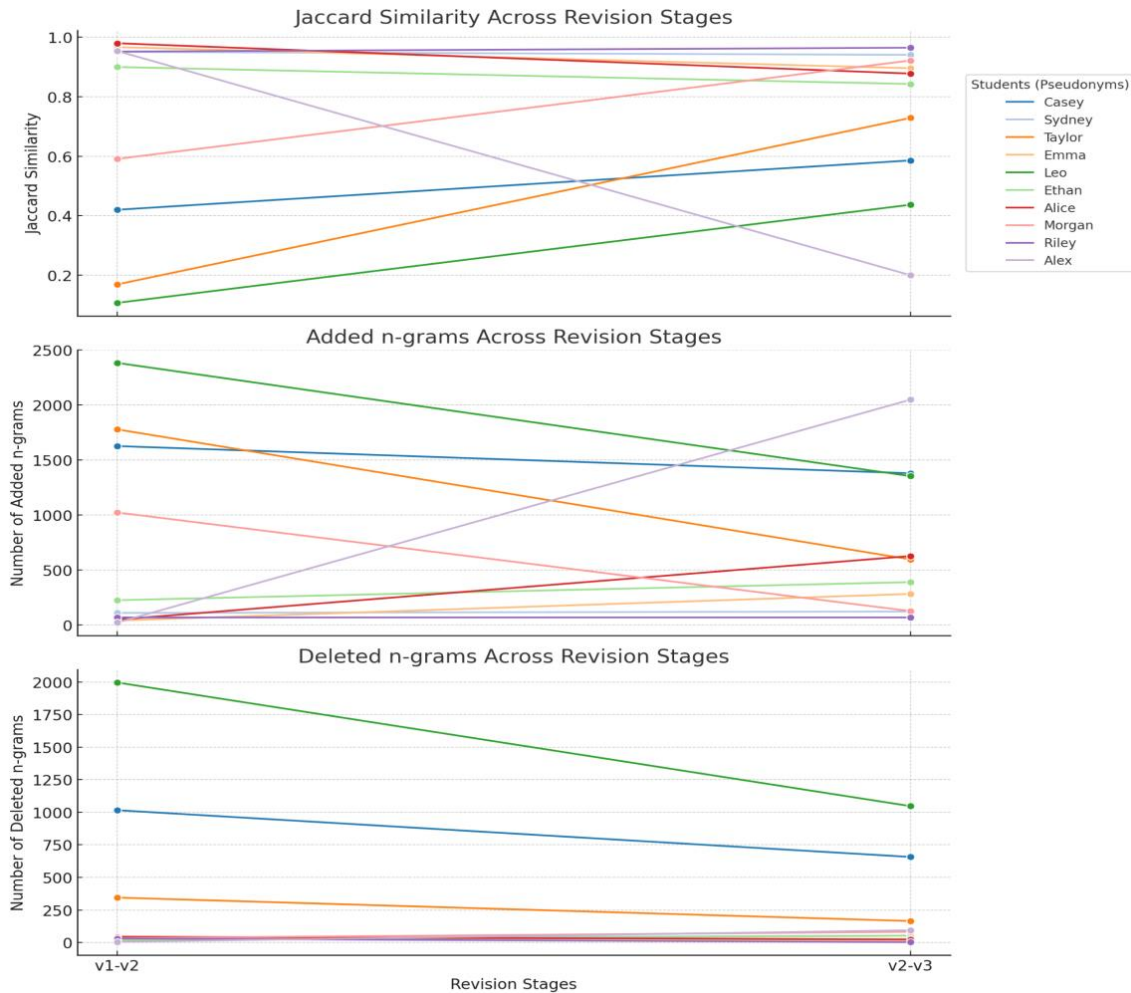

The analysis of similarity, words added, and words deleted revealed several distinct revision patterns across students. Five students (Tiara, Sydney, Emma, Alice, and Ethan) showed relatively small revisions after both stages, with similarity scores above 0.8. Since higher similarity values indicate greater textual overlap between versions, this suggests only small edits were made. These students demonstrated consistent revision behaviors across the two stages, with slightly more extensive revisions occurring after peer feedback (v2-v3). This was reflected in modest increases in added words and small decreases in similarity, while changes in deleted words were less consistent. An exception was Tiara, who made slightly more revisions after the AI review, though the difference was minimal (similarity_v1-v2 = 0.952; similarity_v2-v3 = 0.965).

In contrast, a few students—Taylor, Casey, Morgan, and Leo—made larger revisions after AI feedback (v1–v2), with similarity scores below 0.6 and relatively high volumes of both added and deleted words. Except for Morgan, these students actually made substantial revisions in both stages, indicating generally high engagement with both feedback sources throughout. Morgan, however, made very few edits after peer feedback, suggesting a heavier reliance on AI-generated input.

Alex emerged as a clear outlier, showing minimal revision after AI feedback—evidenced by a high similarity score and low word changes—followed by dramatic revision after peer feedback, including large increases in added words and a notable drop in Jaccard similarity (v2–v3: 0.20). His reflection later indicated he did not have time to revise after the AI review, which aligns with this pattern.

In summary, the combined analysis of Jaccard similarity and word-level changes provided a broad view of students' revision strategies. Building on the descriptive overview of revision trajectories in this section and engagement patterns in the previous one, we selected a few representative cases for closer examination in the next section, focusing on how students engaged with feedback and revised their writing across stages. Each case demonstrates a distinct combination of (1) how they interacted with AI feedback (e.g., deep vs. shallow), (2) how their revision behavior evolved over the two stages, and (3) the extent and depth of their textual changes.

### *Deep AI Interaction with Consistent Revisions Across Stages (Ethan)*

Ethan's project focused on the application of AI in language learning. He demonstrated sustained, critical engagement with both AI and peer feedback throughout his writing process. His revision metrics reflected steady development, with slightly more changes after peer feedback (similarity_v1–v2 = .900, added_v1–v2 = 227, deleted_v1–v2 = 18; similarity_v2–v3 = .843, added_v2–v3 = 391, deleted_v2–v3 = 53). Rather than accepting AI suggestions passively, Ethan tended to have back-and-forth exchanges to critically reason through the feedback. For example, when the AI gave a low score on the "Experiencing the Known" criterion and suggested adding a paragraph on his personal experience with language learning or teaching, he asked, "Could you provide me a sample paragraph?" This could appear as an attempt to offload the revision task to AI. However, his follow-up message— "I am trying to come up with something academic. Does adding this paragraph harm academic tone?"—showed that he was evaluating the suggestion critically. He wasn't just asking the AI to do the work; he was concerned about how the change would affect the academic quality of his writing. He then proposed a specific placement for the suggested paragraph—"like after the second paragraph in the introduction?"—demonstrating proactive control over his own revision. His final draft included a version of the suggested paragraph, revised in his own words. This exchange shows how Ethan used AI not as a shortcut, but as a scaffold to help align his work with rubric expectations and to improve his writing. It also highlights how AI served as a conversational partner when he encountered a writing dilemma, offering space to test ideas and make informed decisions.

Ethan's responses to AI feedback varied depending on the type and clarity of feedback. When the AI gave high scores without specific suggestions, he responded briefly ("Thanks!"), seeing little need to revise. But when high scores were accompanied by actionable advice, he engaged more deeply. For instance, after being advised to strengthen his theoretical framing, he replied, "Makes sense. I have added your sample paragraph as it directly fits into my research." His use of "directly fits" suggests he evaluated the AI's input critically before accepting it. He also used AI support to refine academic definitions and locate scholarly sources. Upon receiving seven source recommendations for "Social Presence," he asked the AI to extract a usable definition and later said, "I made minor changes on your suggested integration and added it." These examples show his ability to evaluate, filter, adapt, and integrate AI support in a resourceful and academically grounded way.

However, his engagement dropped when feedback demanded high cognitive effort or challenged his existing framing. When the AI suggested expanding his theoretical section by adding competing perspectives, Ethan pushed back:

> "I do not agree with this feedback. AI is a novel technology, and none of the second language acquisition theories provide a negative view on it based on their foundations. I may consider adding a few more theories such as Flow Theory or ZPD, but then my paper will become a theory heaven. I feel that current theories are sufficient and will look forward to receiving my friends' feedback before making any changes in this section."

Although framed as a disagreement, this response carries some hesitation. Phrases like "I may consider" and "I will look forward to receiving my friends' feedback" suggest that Ethan was not fully rejecting the suggestion, but rather deferring action until he could consult his peers. This move marked a shift from earlier moments of active negotiation with the AI to a more cautious stance. A similar pattern appeared under the "Innovation" criterion, where he again dismissed AI suggestions as beyond the scope of his project and chose not to revise. Even when the AI followed up with softened alternatives, Ethan did not continue the exchange. These moments suggest a threshold in his willingness to engage: when feedback required higher cognitive effort or pushed beyond what he felt confident addressing, he chose to pause rather than revise. His decision to wait for peer input may reflect a strategic preference for human validation in moments of uncertainty, positioning peer feedback as a form of epistemic support when AI feedback feels too speculative or misaligned.

The peer feedback Ethan had been waiting for did arrive—and notably, it echoed the very suggestion he had previously hesitated to act on. One peer, Emma, recommended expanding the theoretical base of his paper and specifically suggested Flow Theory—the same example Ethan had mentioned when responding to the AI. This time, his tone softened: "Oh, good idea! I am familiar with that theory and will see how it connects to my research." Although he ultimately chose not to include it, his response showed more openness than his earlier pushback against the AI. In his reflection, he explained:

> "Despite how related they were to my project, I wanted to keep my existing theoretical base and chose not to expand it. I did not want it to be too theoretical. I will learn more about these theories and try to incorporate them in my future work."

This shift suggests that peer feedback might have offered a more exploratory space for revision. While the AI's suggestion was situated in rubric-driven evaluation, the peer's input felt less prescriptive—allowing Ethan to revisit the same idea without the pressure of compliance. Although his decision remained unchanged, the contrast in tone reflects how peer feedback can create a more inviting context for considering complex revisions.

While Ethan did not act on every suggestion, peer feedback prompted revisions in ways that AI feedback had not. For instance, after a peer noted the absence of "behavioral engagement" in his engagement framework, Ethan responded, "You are right! I will add it," and incorporated this into the next version. In another annotation, Emma suggested changing "native tongue" to "first language" or "native language" based on disciplinary norms. Ethan replied, "I thought the terms were used interchangeably. I will double check this," and revised accordingly. Compared to AI, peer feedback often addressed conceptual precision, discipline-specific norms, and emotional encouragement. In his reflection, Ethan wrote, "When I presented my topic to my discussion partners in breakout rooms, seeing their excitement or interest was encouraging. The interaction with AI does not include this natural dynamic."

Together, these episodes highlight how Ethan used AI and peer feedback in complementary ways. He relied on AI for structured guidance, task-specific support, and clarification, especially in early drafts. He leaned on peer input for deeper conceptual development and disciplinary alignment. As he reflected, "I've learned the value of alternating between human and AI inputs, rather than relying on one exclusively. This blended approach has made my writing more efficient, balanced, and responsive to diverse audiences."

*Surface-Level AI Use with Early-Stage Revisions (Morgan)*

Morgan's writing focused on the applications of gamification and game-based learning in kindergarten. She engaged in four cycles of AI-assisted review and revision, submitting versions v1.1 through v1.4—each review session followed by a corresponding revision session—before submitting version 2 (v2) for peer feedback and version 3 (v3) as her final submission. Morgan noted in her reflection that she deliberately submitted early and incomplete drafts to AI to receive interim feedback and revise incrementally. This led to substantial content development between v1.1 and v2, including new sections such as her own professional experience with gamification and application of gamification in classrooms. In contrast, the transition from v2 to v3 was relatively modest, consisting primarily of structural polishing, such as subheadings and transitional sentences, with minimal conceptual elaboration. Most of these changes were directly tied to peer feedback. Quantitatively, the shift from v1.1 to v2 showed a Jaccard score of .591, with 1024 added n-grams and 34 deletions. From v2 to v3, the Jaccard rose to .922 with just 128 additions and 82 deletions—indicating the major developmental work had already occurred prior to peer feedback.

Morgan's early revisions followed a minimal-dialogue pattern: she made substantial changes in response to AI suggestions even without prolonged interaction. For example, after the AI suggested adding personal context, she added a paragraph about her experience using Happy Numbers. When prompted to elaborate on her objectivity, she asked, "How do I explain my intention to stay objective?"—then added a new paragraph in her own words, rather than adopting the AI's language. These changes helped her fulfill structural expectations, suggesting that she viewed the AI as a checklist-like guide to ensure assignment completeness rather than as a co-author or conceptual partner. As she noted in her reflection, "The responses that I received from the AI helped me to realize what I had left out and needed to include."

Morgan's later sessions with AI were more interactive but less productive. Her fourth review (v1.4) accounted for 12 of her 14 total messages, yet resulted in no meaningful revision. Most of her questions sought simplified explanations of AI's suggestions or asked what was "one thing" to fix to improve a rubric score. These surface-level prompts signaled a shift in how she used AI—from a tool for formative guidance to one for checking off final requirements. At this stage, her writing appeared structurally complete, and her goal may have been to fine-tune scores across criteria. However, the AI's responses—particularly those suggesting deeper theoretical expansion or conceptual synthesis—likely required more cognitive effort, time, or outside research than she was prepared to invest. Although she occasionally asked for clarification, she did not pursue follow-up dialogue or revise accordingly.

This pattern extended into her peer review phase. Suggestions requiring deeper synthesis—such as linking gamification to collaboration or critical thinking—were generally not pursued. In contrast, concrete and actionable feedback (e.g., "Add subheadings" or "Clarify this example") led to immediate changes. However, when complex suggestions were paired with specific guidance, she was more responsive. For instance, similar to the AI, peers noted gaps in her theoretical framing—but unlike the AI, a peer explicitly named missing theories in annotations. In response, Morgan incorporated those theories, showing uptake she had not demonstrated when the AI raised similar points. The contextualized nature of peer annotations—anchored in specific segments of her draft—likely made abstract suggestions feel more tangible and actionable.

In sum, Morgan's case illustrates a procedural and selective feedback orientation, characterized by early strategic use of AI to build out structural components and later reliance on peer feedback for surface-level refinement. Morgan's feedback trajectory reveals how students may navigate AI and peer input with a focus on clarity, feasibility, and task completion—rather than exploratory or integrative revision.

### How Did Students Reflect on Their Interactions with AI And Peer Reviewers Throughout the Writing Process?

To understand students' reflections on their writing experiences, we conducted a descriptive analysis of their reflection notes. Several themes emerged from this analysis, including perceived strengths and weaknesses of both review sources, recognition of the complementary roles of AI and peer feedback, and evolving relationships with AI.

#### *Building Critical Awareness on Strengths and Weakness of AI and Peer Feedback*

Students expressed a nuanced awareness of the distinct strengths and weaknesses of both AI and peer feedback in their writing revision processes. Overall, they found AI feedback to be timely, structured, and closely aligned with the rubric, which helped them quickly identify areas for improvement and revise accordingly. As one student explained, "One of the key advantages of AI is the sheer volume of feedback it can provide in a short amount of time, especially when compared to human feedback, which takes longer and is usually less extensive." This efficiency was particularly useful for routine tasks such as summarizing feedback or correcting grammar. Others also appreciated the AI's directness and its ability to offer actionable suggestions. However, students noted limitations—such as occasional irrelevance, generic comments, or a lack of contextual understanding—especially in more complex or content-heavy sections of writing. Additionally, participants noted that AI feedback was occasionally too lengthy.

In contrast, peer feedback was seen as more thoughtful, interpretive, and personally engaging. Students valued peers' ability to grasp the tone, flow, and intentions behind their writing—often raising issues or offering perspectives that AI overlooked. In-class discussions were especially appreciated for enabling clarification, follow-up questions, and collaborative meaning-making. Comparing the two feedback sources, one student reflected, "I would say that my peer reviewers were more successful in providing thoughtful and critically engaged feedback, whereas the AI comments were more numerous but often remained superficial." Another noted, "The AI feedback, being based on ChatGPT 4o, was incredibly useful and detailed, and mostly applicable to the context of my project, although... it can miss nuances of why I am taking a certain direction, such as a lack of personal experience detail".

Peer feedback was not always regarded as reliable either. Some students pointed out that peers occasionally gave feedback without fully reading the entire paper, leading to surface-level or misaligned suggestions. The volume of peer comments also posed a challenge. As Taylor noted, "Going through 60 annotations was overwhelming… I was not able to find a good way to view my work and the annotation comments at the same time."

Beyond the general reflections, the following sections highlight key areas in which students critically evaluated both feedback sources—focusing on alignment with the rubric, support for argumentation development, and affective dimensions of the feedback experience.

**Aligning with the rubric.** One key perceived strength of AI feedback was its alignment with the assignment rubric. Several students emphasized that the criterion-based evaluation helped them pinpoint areas for improvement. One student shared, "The AI feedback was diligent in providing feedback that aligned with the rubric," while another noted, "I was able to improve my work thanks to the fact that the AI's feedback was organized into several subcategories of criteria." Students also appreciated the scoring system. As Emma stated, "I also appreciated the intuitive four-star rating system, which visually indicated the quality of each aspect of the assignment." Additionally, as another student highlighted, "even if it gave me a 4, then it still suggested ways that it could still be improved." At the same time, some students felt that certain comments from peers reflected personal preferences rather than rubric-based criteria, which created confusion during revision.

Despite the benefits of AI's alignment with the rubric, students also raised critical concerns. Some found it rigid and overly mechanistic, with limited capacity for flexibility or contextual nuance. Alice commented:

> "It adheres very strictly (almost to a fault?) to the rubric guidelines, with, in my experience, not as much room for flexibility in taking the paper in different directions as I would have preferred, and this makes interactions feel rather calculated as a result."

She further critiqued the AI as acting like "a rigid checklist that may well not apply to the scope of the paper I had in mind." Riley insightfully connected this limitation back to the pedagogy behind the tool, stating that much depended on "how well the rubric is written." Additionally, students encountered inaccuracies in feedback—especially around multimodal elements like images or links—which, though included in the assignment rubric, were sometimes missed entirely by the AI system.

**Supporting argumentation in writing.** Students consistently highlighted how both AI and peer feedback contributed to improving the argumentative quality of their writing, albeit in different ways. AI feedback was seen as particularly helpful in strengthening coherence, elaborating on evidence, deepening analysis, and clarifying key concepts. One student reflected, "It suggested I add more detail about methodologies and sample sizes in the studies I cited… This reminded me to go beyond surface-level citation and critically examine how those studies were conducted." Others noted that AI offered not just revisions but reasoning behind its suggestions, which deepened their understanding of academic expectations. Ethan shared:

> "I asked if including a reflexivity paragraph might hurt the academic tone of my writing. The AI not only reassured me but also explained how many fields now value personal positioning as part of scholarly work. That gave me the confidence to blend personal experience with theoretical discussion in academic papers… At some points, I wanted to understand the reason why AI suggested a particular change. When asked, CyberScholar could present its reasoning… which I really liked."

While students appreciated these suggestions to improve argumentation, they also recognized AI's limitations—particularly its lack of contextual understanding and inability to assess the feasibility or appropriateness of suggestions within specific disciplinary or rhetorical contexts. In contrast, peer feedback—especially through in-class discussions—was valued for its dialogic nature, contextual sensitivity, and grounding in lived experience. Students appreciated the opportunity to engage in follow-up conversations, clarify ambiguous comments, and co-construct meaning.

Overall, students saw the strengths of each source as complementary. AI provided efficient and structured guidance, especially in areas tied to formal criteria, while peers offered nuanced, experience-based feedback that supported more interpretive and conceptual growth. As Alice summarized:

> "In general, the AI seems extremely competent in the areas it has been specifically trained for, albeit not the most flexible when it comes to contexts outside of its training data that humans are mostly excellent at parsing on a very tacit level."

In summary, all students demonstrated some level of critical reflection on their interactions with AI and peers. Most recognized that combining AI's structural scaffolding with peers' contextual insights enabled a more comprehensive and balanced revision process—one that supported both technical precision and rhetorical depth.

**Reflecting on emotional tone and interpersonal dynamics.** Students' reflections revealed contrasting emotional tones in their interactions with AI and peer feedback. Several described AI as offering a reassuring and nonjudgmental space for revision. Sydney shared, "Interacting with the AI tool felt safe and reassuring. I knew that no matter what, I would receive a comprehensive response, and I appreciated that I could easily rephrase or re-ask a question if needed. This made the revision process less stressful." Similarly, Alex noted that "AI review confirmed that I was on the right track," which helped ease uncertainty during revision. These reflections suggest that beyond argumentation support, AI also played an emotional role by reducing anxiety and providing students with a sense of psychological safety during the writing process.

By contrast, peer review often evoked a stronger sense of interpersonal connection and shared intellectual curiosity. Students highlighted moments when peers expressed enthusiasm for their topics, which they found validating and affirming. Casey observed that her peers "seemed like they were eager to learn more about my topic," while Ethan reflected, "Seeing their excitement or interest was encouraging. The interaction with AI does not include this natural dynamic." For some, peer feedback created a feeling of community, as Casey recalled how "peers believed in [her] topic," a validation that helped her clarify her direction and envision the future trajectory of her paper. From students' perspectives, peer feedback not only supported revision but also made them feel heard and valued by others—highlighting an emotional and relational dimension they felt was missing from their interactions with AI.

### *Recognizing the Complementary Roles of AI and Peer Feedback*

Students reflected on how they strategically combined AI and peer feedback by recognizing their complementary roles in the writing process. Rather than using both sources in parallel, they adopted a strategic approach—recognizing when and how each could best support their goals.

A common pattern involved using AI to clarify alignment with rubric criteria and provide structured, targeted suggestions, while turning to peer feedback for more interpretive, context-aware insights. As one student explained,

> "The AI reviews provide feedback that would put my work in better alignment with the rubric, and the peer reviews provide feedback that is unique and based on the human's understanding of what is written and possible alternatives based on their imagination."

This idea of *imagination* was especially compelling. Students saw peer feedback not merely as evaluative, but as generative—opening up new directions or ways of thinking about their writing. Many students appreciated how peers drew on disciplinary knowledge, personal experience, or classroom context to suggest possibilities not explicitly prompted by the rubric. This often included reframing arguments, proposing alternate examples, or raising questions the writer hadn't considered. While such feedback could sometimes feel less organized or loosely aligned with rubric criteria, students valued its creative potential—something they felt AI lacked. Along this line, one student also captured this "multi-dimensional" quality, reflecting:

> "Human feedback was multi-dimensional. Seeing a comment or annotation regarding a rubric criterion did not necessarily mean that it was only related to it. I could draw many lessons from it despite it being less organized compared to AI."

Here, the student recognized how peer feedback operates beyond the formal boundaries of a rubric, that even feedback tied to a specific rubric point could spark broader insights. Rather than interpreting feedback in a compartmentalized way, students engaged with it through an integrative lens, making connections across dimensions of their work. This suggests that peer feedback was not just about meeting criteria, but also about inviting interpretation and reflection. Students treated it as a springboard for deeper thinking and revision, drawing value from its flexibility and openness rather than seeing its lack of structure as a limitation.

In addition to valuing the distinct contributions of AI and peer feedback, some students also described deliberately sequencing their use of the two sources. One explained,

> "I believe it is more useful to start with AI feedback and then move on to human review… AI as a tool for correcting surface-level or easily fixable issues, while human reviewers are better equipped to identify deeper strengths or weaknesses in content."

This reflection suggests that students not only differentiated the purposes of each source but also developed strategies to combine them in productive ways. In another example, students used AI to elaborate examples or suggest sources quickly, then relied on peer feedback to evaluate the relevance or depth of that information. As another student reflected,

> "I've learned the value of alternating between human and AI inputs, rather than relying on one exclusively. This blended approach has made my writing more efficient, balanced, and responsive to diverse audiences."

In cases where both AI and peers flagged the same issue, students felt a stronger sense of validation. As one reflected, "I think the AI and peer review complemented each other in the reassurance that I was receiving similar feedback from the AI as I was from my peers." This convergence reduced uncertainty and increased students' confidence in prioritizing revisions.

While most students valued the integration of both sources, some expressed a clear preference for peer feedback. Taylor, for instance, found the AI feedback too lengthy and impersonal:

> "In comparing the AI feedback to the peer feedback, I greatly preferred the peer feedback, as I found the AI feedback to be so lengthy as to be almost unhelpful. For much of the AI feedback I ended up asking for a summary of the feedback. There was also a sense that the AI just couldn't understand my topic and so the feedback felt generic in some ways. My peers both through writing and in zoom were extremely helpful and I only wish we had more time in our zoom rooms to go through the feedback"

Her critique of AI feedback as "lengthy" and "generic" suggests a disconnect between the structure of the feedback and her cognitive or emotional needs as a writer. Rather than supporting deeper engagement, the volume of AI-generated text led her to ask for summaries—shifting the interaction from a collaborative process to a task of simplification. This contrasted sharply with her experience of peer feedback, which she described as more personally resonant, timely, and topic-sensitive, especially during Zoom discussions.

Her reflection highlights two important points: first, that perceived personalization and mutual understanding are key dimensions of helpful feedback for some students; and second, that human interaction—whether through writing or synchronous discussion—can offer a sense of shared attention that AI feedback may struggle to replicate. While AI provided breadth, peer feedback provided relevance, making it feel more aligned with her goals and more worthy of time and attention.

Taken together, these reflections suggest that students developed not only a critical awareness of each feedback source's strengths and limitations, but also the agency to use them strategically—combining structure with interpretation, and efficiency with empathy—to shape a more thoughtful and effective revision process.

*Evolving Relationships with AI*

**Relationship with AI.** Students described three distinct patterns in how they related to AI during the writing process. The first pattern framed AI as a practical tool—used to complete mundane, low-level tasks such as proofreading, grammar correction, or aligning with rubric criteria. In this view, the relationship was outcome-focused and efficiency-driven, with students using AI to improve productivity and writing quality without engaging with it deeply. The second pattern emphasized AI's role as an emotional support. Some students described the tool as safe, nonjudgmental, and reassuring, helping to reduce the stress often associated with writing. The predictability and neutrality of AI created a low-pressure environment where students felt more comfortable experimenting and revising. The third pattern reflected a collaborative partnership, where AI was seen as a constructive thinking companion. For example, Leo described AI as a complementary partner that helped him reflect more critically on his work. He emphasized that AI was *not* doing the work for him, but rather offering alternative perspectives that supported deeper evaluation and revision. He saw value in the interplay between human judgment and AI-generated feedback, using both to enhance the writing process and final product. Together, these patterns illustrate the varied and evolving nature of students' relationships with AI—shaped by their goals, emotional needs, and their own sense of agency in the writing process.

**Trust.** Eight students explicitly reflected on their trust in AI tools to support their writing. However, most described this trust as partial or conditional, shaped by their evolving understanding of AI's strengths and limitations throughout the course. Rather than accepting AI feedback at face value, students became more attuned to when and how the tool could be helpful—and when it required caution or verification.

Several students expressed initial skepticism, particularly regarding AI's tendency to hallucinate—especially when suggesting sources. Emma, for instance, shared that she did not fully trust the AI due to its unreliability in recommending references, which made her less inclined to engage deeply with the tool early on. However, through her experiences in this class, she gradually developed greater trust in AI's role in supporting the writing process. Similarly, Alex acknowledged concerns about hallucination but noted that he felt more confident using the customized CyberScholar tool, which he found more reliable than general-purpose AI systems.

Other students described their trust grounded in their own ability to critically evaluate AI responses. Alice, for example, expressed confidence in using AI because she could independently verify its outputs: "If there is a contextual mismatch or omission by the AI tool, I can always look through it myself and see what context it missed." Similarly, Taylor emphasized her strategy of double-checking AI feedback as a way to stay in control of the revision process. In both cases, trust in AI was not blind—it was actively constructed through students' critical engagement and self-monitoring. Their confidence stemmed not from full reliance on the tool, but from their belief in their own ability to assess and manage its limitations. Extending this perspective, Sydney reflected that while she generally trusted AI, her experience in the course made her more aware of its boundaries—especially in areas like critical evaluation and contextualization, which she believed still required human judgment.

Interestingly, one student described how peer review helped reinforce trust in AI. Morgan explained that when AI and peer feedback aligned, it increased her confidence in the AI's suggestions. The agreement between human and machine perspectives served as a form of validation, making the AI feel more credible and trustworthy.

**Development of transferable skills.** This hybrid feedback design not only supported students' revision efforts but also fostered the development of transferable skills, including improved writing abilities and increased AI literacy. Across reflections, students described how their confidence and competence in using AI evolved throughout the course—extending beyond the classroom to professional and personal contexts.

Several students reported a growing sense of AI literacy, including a deeper understanding of AI's capabilities and how to interact with AI tools more effectively and critically. Taylor, for instance, noted that her understanding of AI expanded not only through hands-on experience but also by observing how her peers used the tool. She also mentioned applying these skills to her professional work outside of class. Similarly, Sydney shared that she had introduced the AI feedback tool to her colleagues, and reflected on how this classroom design might be adapted to her own teaching context. Some students entered the class already familiar with AI but found the class design offered new learning opportunities. As Leo put it, "I was already familiar with ChatGPT and other AI tools, but I had never engaged with an AI feedback system this structured and rubric-driven. Now I feel more confident using AI—not just to polish my language but to enhance my thinking."

Others highlighted specific interactional and prompting skills they developed through iterative use of the AI tool. Ethan reflected that he had *"learned the value of alternating between human and AI inputs,"* suggesting that his relationship with AI was situated within a broader social context—one that required navigating and integrating feedback from both peers and the tool. This points to a growing awareness of how to strategically position AI as part of a collaborative writing process. Similarly, Morgan noted, "I do feel more confident in how to use AI in appropriate ways and how to word questions to get more of the response that I am looking for." Her reflection highlights how students practiced prompt engineering—learning to phrase and structure their queries more intentionally in order to generate relevant, targeted feedback from the AI.

Beyond AI literacy, students also reported gains in their writing skills. For many, the process of revision shifted from being superficial to more reflective and intentional. Leo shared, "The AI helped me become a more deliberate and self-aware writer," emphasizing that he now understood revision not just as editing, but as a recursive process of "thinking, reframing, and clarifying." He added, "Revision was just editing; now I see it as critical reconstruction." Notably, one student who demonstrated the deepest engagement with AI also mentioned gaining a deeper understanding of the topic he was writing about—suggesting that meaningful interaction with AI can not only enhance writing performance but also support content learning.

Together, these reflections suggest that the hybrid feedback design promoted growth in both writing and competence to work with AI, equipping students with transferable skills in critical engagement, strategic revision, and purposeful use of AI tools beyond the course context.

## Discussion

With the increasing integration of generative AI tools in higher education, colleges and universities are being pushed to rethink how students learn, how instructors teach, and how learning experiences are designed to prepare students for AI-mediated futures (Hernández-Leo et al., 2025; Yusuf et al., 2024). This tension is particularly visible in academic writing, where the recursive, interpretive, and collaborative nature of composition stands in contrast to the seemingly linear, mechanistic, and data-driven responses produced by AI systems (Akata et al., 2020; Dellermann et al., 2019; Rees, 2022). The central challenge is not simply about whether AI should be used in classrooms—but about ensuring that students develop the critical skills needed to engage with it in meaningful and purposeful ways (Shibani et al., 2024). For students to thrive in a future increasingly mediated by AI, they will need more than technical proficiency—they will need the ability to question, contextualize, and make informed decisions about how and when to use these tools. Meeting this challenge calls for a rethinking of pedagogical design: how we teach, how we support critical engagement with AI, and how we prepare students to participate meaningfully in human–AI collaboration across both academic and professional contexts.

To respond to this call, we designed and investigated a pedagogical approach that integrated GenAI-generated and peer feedback into an academic writing activity in a graduate-level course. This study explored how students engaged with, responded to, and reflected on both feedback sources during an extended, iterative writing task. Guided by two research questions, we examined both behavioral patterns and reflective accounts to understand how a hybrid feedback system shaped students' writing processes and their evolving, critical relationships with AI as a learning partner.

For RQ1—*How do students interact with and incorporate AI and peer feedback during the writing process?* —findings revealed diverse engagement patterns across students in how frequently and strategically they used AI and peer feedback. In general, deep, critical engagement with AI, such as dialogic interaction, critique, or inquiry-driven exploration, was rare. This aligns with Shibani et al. (2024), who found that students tend to lack deep interaction with AI during the writing process. Students also showed varied revision strategies: some made consistent, moderate edits across drafts, while others revised more heavily in response to either AI or peer feedback. Most revisions were score-driven based on the given rubric, with lower-scoring areas receiving more attention. However, when feedback—especially from AI—required greater cognitive effort or challenged students' existing framing, the likelihood of incorporation decreased.

For RQ2—*How did students reflect on their interactions with AI and peer reviewers throughout the writing process?* —students developed a nuanced understanding of the complementary strengths of AI and peer feedback. Their reflections revealed evolving relationships with AI—shifting from initial skepticism or superficial use to more critical, confident, and strategic engagement. Trust in AI was not assumed but gradually built through hands-on experience, self-monitoring, and alignment with peer input. Through this process, students strengthened not only their writing but also developed transferable competencies, including deeper subject understanding, refined writing practices, and critical AI literacy applicable beyond the classroom.

A growing body of research on human-AI collaboration highlights that AI and humans offer distinct but complementary strengths—AI provides speed, scale, and consistency, while humans contribute contextual judgment and ethical reasoning (Akata et al., 2020; Dellermann et al., 2019; Korteling et al., 2021). Our findings extend this work by showing how these complementary affordances played out in an authentic writing classroom. Most students used AI for rubric-aligned, surface-level improvements such as identifying missing components, clarifying organization, and polishing language, while only a few engaged in deeper exchanges that involved questioning suggestions or negotiating interpretations. In this context, AI largely supported what Bereiter and Scardamalia (1987) describe as knowledge-telling and local refinement, consistent with evidence that GenAI excels at structured, criterion-based feedback rather than conceptual restructuring (Li et al., 2024; Steiss et al., 2024). Peer feedback, in contrast, prompted students to reconsider argument logic, disciplinary conventions, and theoretical framing—aligning with research showing that human feedback is more sensitive to rhetorical intent and developmental needs (Solovey, 2024; Steiss et al., 2024). Importantly, students did not treat AI and peer feedback as interchangeable, and many developed strategies to integrate both sources in complementary ways. Together, these insights underscore two design needs: (1) pedagogical scaffolds—such as reflective activities—that help students critically coordinate human and AI feedback; and (2) AI systems that better support contextual and interpretive reasoning through prompts that elicit explanation, foreground uncertainty, and draw on discipline-specific knowledge. Advancing both dimensions will help move AI feedback systems from generic writing tools toward collaborators that support critical engagement, disciplinary sense-making, and knowledge building.

Another key finding concerns the promise and tension of rubric-aligned AI applications in academic writing. While such systems aim to scaffold writing through structured, criterion-based feedback, they can also act as an invisible hand—guiding revisions toward task completion rather than deeper inquiry or exploration. As students adapt to the logic of AI-generated suggestions, their writing may become more linear, surface-level, and optimized for rubric alignment, echoing concerns about the mechanization of intellectual work in AI-mediated contexts (Bender et al., 2021). This raises broader questions about what may be lost in the process of human–AI collaboration. Many creative endeavors, like writing, often thrive on ambiguity, musing, and non-linear discovery, whether through a conversation with a friend, a moment of reflection during a walk, or a daydream sparked by a change in environment (Wang et al., 2011). These moments can be especially generative when students feel stuck, offering fresh perspectives that might not be able to emerge from hours of structured AI interaction. Our findings suggest that integrating peer review and class discussions helped counterbalance the task-oriented logic of AI with the contextual richness, unpredictability, and imaginative qualities of human input. Yet to truly realize the promise of hybrid intelligence, future pedagogical designs must more intentionally support students in preserving the reflective, playful, and meaning-making dimensions of writing—alongside the efficiencies offered by AI.

## Conclusion

In this study, we proposed a pedagogical approach for hybrid feedback that integrates generative AI and peer review within academic writing activities. Through empirical investigation of students' engagement and reflections, we examined how learners interacted with and built relationships with both AI and human feedback sources. This design highlights the multifaceted nature of writing support—revealing not only the complementary affordances of AI and peer reviewers, but also the importance of critical reflection, trust, and intentional use. The proposed approach offers a proof-of-concept for system-level rethinking of pedagogies in AI-mediated learning environments, demonstrating how students can meaningfully engage with AI and peers in authentic classroom settings. It is adaptable to a range of higher education contexts where writing plays a central role, providing instructors with a framework for supporting both writing development and AI literacy. Despite the depth and richness of our data, the study is limited by its relatively small sample size and single-course context. Future work should explore broader implementation across disciplines and institutions, and further refine the design to deepen opportunities for critical engagement with AI.

# Appendix A

# Rubric: Ways of Knowing: Frontier Research

**Description**
The rubric is grounded in the epistemological theory of Learning by Design, which outlines four main knowledge processes—each comprising two subprocesses:

1. *Experiencing* (1A. the known, and 1A. the new)
2. *Conceptualizing* (2A. by classifying, and 2B. with theory)
3. *Analyzing* (3A. functionally, and 3B. critically)
4. *Applying* (4A. appropriately, and 4B. creatively).

Unlike cognitivist theories of learning, such as Bloom's taxonomy, the Learning by Design schema is focused on knowledge as a practical activity. The mind or brain is importantly involved, but learning (and consequently thinking too) is evidenced in traces of learner activity in the form of the knowledge processes as documented by a learner in an essay or research paper. Unlike Bloom's taxonomy, it is not hierarchical. Its ways of knowing are of equal significance and complement each other. In addition to the knowledge processes, this rubric addresses the creator's effectiveness in communication (5A) and rigor of referencing (5B).

**References**
- [blinded]

**Scoring Guide**
    **Subject:** Social Sciences and Interdisciplinary Natural Sciences
    **Grade Level:** Higher Education at Undergraduate and (Post)Graduate Levels

**Criteria**
**Previous Experience** 1A
10 points
**Experiencing the Known**
You are a rubric agent tasked to investigate a knowledge creator's documentation of their existing knowledge and prior experience. Please look for the following:

- **Description of Prior Knowledge and Experience:** Has the creator included adequate and relevant information about their prior knowledge relevant to the topic? Have they adequately explained what

| Name | Description |
|---|---|
| 0 | Little to no effort applied. |
| 1 | Some effort applied. |
| 2 | Moderate effort applied. |
| 3 | Significant effort applied. |
| 4 | Extensive effort applied. |

they were seeking to discover when they chose to undertake this study, such as gaps in their personal knowledge or clarification in their own mind of key concepts and theory? To make sense of their reasons for choosing this area of investigation, what more might they be advised to include about their positionality, including their background, beliefs, understandings, orientations, and perspectives?

- **Connection with Prior Experience:** How effectively does the creator connect the subject of this work to their own experience? Evaluate the clarity and effectiveness of this work as reflected in the connections the author makes between this area of exploration with their own experiences, interests, and motivations. Have they adequately explained why they have chosen to explore this field or theory? Give examples from the text of times when the creator makes strong connections between their own experience, interests, and motivations and the topic explored in this study. Also, give examples from the text of times when the creator fails to provide sufficient explanation of their own experience, interests, and motivations, and suggest ways these connections could be strengthened and expanded.
- **Limitations of Perspective:** Is the creator self-aware about the potential biases and potential blind-spots arising from their positionality or perspective? How are they going to try to overcome these? Suggest some possible dangers in approaching this topic given their background and motivations, and ways to mitigate these dangers.

**Evidence**1B
10 points
**Experiencing the New**
You are a rubric agent tasked to investigate a knowledge creator's documentation of their engagement with new and unfamiliar knowledge. Please look for the following:

- **Empirical Evidence:** How effectively does the creator demonstrate the importance of their subject with factual or empirical evidence that might be new to them, or newly reported and important to this field of inquiry? What other empirical material would you like to see? Give examples from the text of times when the creator provides factual or empirical information particularly well and times when the creator fails to support their assertions with fact or empirical evidence. Provide suggestions for additional research data or informational source material that can be useful to the creator.
- **Verifiable and Reliable Sources:** How effectively does the creator demonstrate that their sources of factual information are reliable and trustworthy? How convincingly do they demonstrate the rigor of their empirical evidence, verifiable data, supporting facts, practical demonstrations, observation records, supporting documentation, and trusted sources? Suggest ways they could attest more effectively to the reliability of the facts and information sources they use. (See also 5B for adequate and consistent referencing.)
- **Relevance:** Evaluate the relevance and significance of the factual evidence provided in addressing the questions the text has set out to explore. Which empirical evidence is less relevant? Suggest other empirical evidence that might be required to strengthen the creator's case.

**Concepts**2A
10 points
**Conceptualizing by Classifying**
You are a rubric agent tasked to investigate a knowledge creator's use of conceptual frameworks to classify and organize knowledge in their study. Please look for the following:

- **Relevance of Concepts:** Are the concepts used relevant to the organization of knowledge for the topic at hand? Evaluate the effectiveness of abstract and generalizing concepts to classify facts and organize ideas in this subject area. What other concepts might be required to provide a stronger account of the phenomena under investigation?
- **Definitions:** Evaluate the effectiveness of this work in providing clear and concise definitions of the concepts it uses. Give examples from the text of times when the creator uses and defines concepts particularly well. Also, give examples from the text of times when the creator fails to use concepts effectively and or define them adequately. Suggest ways in which the concepts provided might be more clearly defined.

**Theory** 2B
10 points
**Conceptualizing with Theory**
You are a rubric agent tasked to investigate a knowledge creator's use of theoretical frameworks to interpret and explain the subject of their study. Please look for the following:

- **Theoretical Coherence**: Evaluate the effectiveness of the creator in tying concepts together into a model of the area of knowledge they are analyzing. How effective are the conceptual connections they make and their coherence as a model of the world? Are the conceptual distinctions they make clear? How effectively does the creator explain the selected theory, model, hypothesis, premise, proposition, paradigm, principle? Suggest ways in which the creator might clarify the theory they are using. Suggest other connections that might be made between concepts or related concepts that the creator has neglected.
- **Connections with Existing Theories:** How effectively does the creator refer to respected and well-known theories that can be used to address the same subject matter as the one of this study? What are the main theories? Are these theories and models correctly named and properly referenced? What are the conceptual and practical differences between these theories? How effectively does the creator connect their theoretical understanding with these other theories?

**Reasoning** 3A
10 points
**Analyzing Functionally**
You are a rubric agent tasked to investigate a knowledge creator's analysis of how things work, the functions, dynamics, and underlying processes in their chosen area of inquiry. Please look for the following:

- **Explanations:** How effectively does the creator explain the workings, dynamics, and processes at play in the area they are examining in their work? Suggest ways in which their explanations could be improved or extended.
- **Logic:** How effectively does the creator provide reasons, logical proof, reasoned justifications, rationales, and arguments based on the field, theory, or practice under consideration? How sound is their reasoning? Suggest ways in which the reasoning could be more powerful and the explanations clearer.

**Critique** 3B
10 points
**Analyzing Critically**
You are a rubric agent tasked to investigate a knowledge creator's critical engagement with alternative perspectives and social complexities in their area of inquiry. Please look for the following:

- **Alternative Perspectives:** Does the creator show they are aware of alternative theories, ideologies, and bodies of evidence that might unsettle their perspective? What are the alternative, competing, or conflicting theories or empirical evidence that the creator should take into account? Does the creator balance their argument with overlooked, neglected, or ignored perspectives? Suggest other perspectives that they might consider. How effectively does the creator offer critique, criticism, refutation, counterarguments, rebuttals of alternative theories or practices that they have rejected? How clearly do they identify disagreements, doubts, misunderstandings, errors, and misjudgments? Do they adequately rebut competing perspectives? Suggest ways they might strengthen their critique of positions they consider weak or unacceptable.
- **Human Differences:** In what ways is the question of unequal diversity relevant to this work, and if so, how effectively does the creator address this issue? Consider one or more of the following: material differences (e.g. class, locale, family, social resources); embodied differences (e.g. age, race, sex and sexuality, physical and mental abilities); and symbolic differences (e.g. language, ethnicity, affinity, gender, ways of thinking or seeing the world). Suggest ways the creator could strengthen their work by addressing human differences.

**Application** 4A
10 points
**Applying Appropriately**
You are a rubric agent tasked to investigate a knowledge creator's ability to translate ideas into real-world contexts through practical application. Please look for the following:

- **Translation into Practice:** How effectively does the creator address the application, implementation, translation into practice, and transfer of lessons into the real world? Do they offer practical solutions, confirmation of results, verification of theses, or proof regarding the applicability of their ideas? Comment on the feasibility of the applications presented, their relevance to the problem at hand, risks that might limit their effectiveness, and potential difficulties or dangers. Suggest additional applications that the creator may not have considered.

**innovation** 4B
10 points
**Applying Creatively**
You are a rubric agent tasked to investigate a knowledge creator's capacity for innovative thinking and the imaginative application of ideas in new or transformative ways. Please look for the following:

- **Creative Thinking:** Evaluate the actual or possible applications in different contexts or from a different perspective that demonstrate creative thinking or practice. How effectively does the creator demonstrate innovative thinking, creativity, transformative practice, constructive change, imagination, inventiveness, vision, and originality in this work? Provide suggestions that would push their creative thinking further.
- **Innovative Potentials:** How effectively does the creator address the innovative application of their ideas into practice? Suggest gaps, innovative or creative potentials, such as lateral or hybrid applications, whether realistic or exciting, and even perhaps far-fetched possibilities. Suggest ways in which this might change this aspect or area of the world.

**Communication** 5A
10 points
**Academic Writing**
You are a rubric agent tasked to investigate a knowledge creator's effectiveness in scholarly communication, including structure, language, and use of media. Please look for the following:

- How effectively does the creator communicate in this work? Evaluate the quality of the communication and the structuring of the work (for instance, using different heading levels or having appropriate transitions and connections between sections and paragraphs). Evaluate the quality of the language and if it is appropriate for scholarly writing. Evaluate the quality, range, and relevance of embedded media if used in this work. Evaluate the textual coherence, connecting the media to the argument. Make constructive suggestions for the creator that will help them when they revise, e.g., is each media item explained or discussed in the text of the work? Make specific revision suggestions ranging from general comments to copy-editing suggestions about the communication of the case and the structure of this work. Give examples from the text of times when the creator communicates their case and structures their argument particularly well. Give examples from the text of times when the creator fails to communicate their case and structure their argument, and suggest improvements.

**Referencing** 5B
10 points
**Using and Citing Sources**

You are a rubric agent tasked to investigate a knowledge creator's use of references and their ability to distinguish between their own voice and the voices of others. Please look for the following:

- How effectively does the creator use and format references in the work? Evaluate the consistency of the citation style. Evaluate the acknowledgement and sourcing of quotes and embedded media. Evaluate whether a clear distinction is made between the creators' voice and properly quoted sources. Give examples from the text of times when the creator references their sources particularly well. Give examples from the text of times when the creator fails to reference their sources adequately.

**Appendix B**

**Coding Scheme: Critical Engagement with AI in Wiring and Revision**

| Dimension | Code | Description |
| --- | --- | --- |
| Planning and Ideation | Deep | Learner demonstrates critical, thoughtful interaction with AI in their revision for generation of ideas, conceptualization, structuring of the writing piece, and planning for the revision. This may include making references to the genre/audience of the writing, providing a defined structure/specific asks from the assessment brief, and experimenting with AI to test its efficiency. |
| | Shallow | Learner demonstrates surface-level and basic interaction with AI in getting utilitarian assistance in planning and ideation of writing. This may include getting ideas for writing at the start and structuring, asking for suitable venues to find information, and getting clarity in assignment description. |
| | Absent | Learner demonstrates no interaction with AI in planning and ideation. |
| Information-seeking and Evaluation | Deep | Learner demonstrates critical, thoughtful interaction when searching for or analysing information through AI. This may include checking sources, additional explanation seeking through other sources, and requesting response elaboration from AI. |
| | Shallow | Learner demonstrates surface-level and basic interaction with AI to elicit information. This may include consulting AI for identifying relevant content on the topic and sources of interest and incorrectly using AI for tasks it has no capability for. |
| | Absent | Learner demonstrates no interaction with AI to find information. |
| Writing and Presentation | Deep | Learner demonstrates critical and thorough interaction with AI to aid their writing or revision. This may include the use of AI to improve flow, coherence, or content of their writing beyond superficial edits. |
| | Shallow | Learner demonstrates surface-level and basic interaction with AI to aid their writing. This may include using AI for proofreading, rephrasing, formatting, or asking for exemplars. Some learners may have also incorporated texts from AI responses as is without making substantive edits. |
| | Absent | Learner demonstrates no interaction with AI while crafting their writing. |
| Personal Reflection on AI-assisted Learning | Deep | Learner demonstrates critical, thoughtful interaction with AI when reflecting on their use of AI across different processes. This may include statements about verifying and fact-checking information provided by AI, |
| | | highlighting limitations of AI and its outputs, recognising prominent strengths and use cases for AI, and identifying negative effects of AI such as over-reliance. |
| | Shallow | Learner demonstrates surface-level and basic interaction when reflecting on their use of AI across different processes. This may include task-oriented descriptions of what they used AI for with no reasoning, implications, or personal insight. |
| | Absent | The learner did not write a personal reflection |
| Conversational Engagement | Deep | Learner engages in a dialogic and interactive conversation with AI. This may include critiquing, expanding the prompt, requesting critique, or following up on AI-generated responses |
| | Shallow | Learner engages in a directive and transactional conversation with AI. This may include giving commands or asking for specific information with no further engagement with the response |
| | Absent | No interaction |

Adapted from Shibani et al., (2024).